\documentclass[lettersize,journal]{IEEEtran}
\IEEEoverridecommandlockouts

\usepackage{amsmath,amsfonts}
\usepackage{algorithmic}
\usepackage{array}
\usepackage[caption=false,font=normalsize,labelfont=sf,textfont=sf]{subfig}
\usepackage{textcomp}
\usepackage{stfloats}
\usepackage{url}
\usepackage{verbatim}
\usepackage{graphicx}
\def\BibTeX{{\rm B\kern-.05em{\sc i\kern-.025em b}\kern-.08em
    T\kern-.1667em\lower.7ex\hbox{E}\kern-.125emX}}
\usepackage{balance}
\usepackage{caption}
\usepackage{siunitx}
\usepackage{tikz}
\usetikzlibrary{shapes.geometric, arrows.meta, positioning, calc, fit}

\usepackage{algorithm}

\usepackage{mathtools}	
\usepackage{amssymb}
\usepackage{amsthm}

\usepackage[utf8]{inputenc}

\usepackage{ifthen}

\usepackage{microtype}

\usepackage{caption}

\usepackage{bm}

\usepackage[draft,nomargin,marginclue,inline]{fixme}
\makeatletter
\renewcommand*\FXLayoutMarginClue[3]{%
  \marginpar[%
  \raggedleft\@fxuseface{margin}\textcolor{red}{\ignorespaces $ \Rightarrow $}]{%
    \raggedright\@fxuseface{margin}\textcolor{red}{\ignorespaces $ \Leftarrow $}}}
\makeatother	

\usepackage{tumcolor}

\usepackage{pgfplotstable}	

\pgfplotsset{
	discard if/.style 2 args={
        x filter/.append code={
            \edef\tempa{\thisrow{#1}}
            \edef\tempb{#2}
            \ifx\tempa\tempb
                
            \fi
        }
    },
    discard if not/.style 2 args={
        x filter/.append code={
            \edef\tempa{\thisrow{#1}}
            \edef\tempb{#2}
            \ifx\tempa\tempb
            \else
                
            \fi
        }
    }
}

\usepackage{cite}

\usepackage[acronym,shortcuts]{glossaries}
\newacronym{cnn}{CNN}{convolutional neural network}
\newacronym{ula}{ULA}{uniform linear array}

\usepackage{cleveref}

\tikzset{algorithm1/.style={mark options={solid},color=TUMBeamerBlue, line width=\lineWidth, mark=square, dashed}}

\pgfplotscreateplotcyclelist{miko}{%
{color=TUMBeamerRed, dash pattern=on 5pt off 1pt on 1pt off 1pt, line width=1.5pt},%
{color=TUMBeamerOrange, dash pattern=on 5pt off 2pt, line width=1.5pt,
mark=o},%
{color=black, dash pattern=on 5pt off 2pt on 3pt off 2pt, line width=1.5pt},%
{color=TUMDarkerBlue, line width=1.5pt},%
{color=black, dotted, line width=1.5pt},%
{color=TUMBeamerGreen, dotted, line width=1.5pt},%
{color=TUMBeamerYellow, dotted, line width=1.5pt},%
{color=TUMBeamerDarkRed, dotted, line width=1.5pt},%
{color=TUMBeamerLightGreen, dotted, line width=1.5pt},%
{color=TUMIvory, dotted, line width=1.5pt},%
{color=TUMLightBlue, dotted, line width=1.5pt},%
}

\DeclareMathOperator*{\argmax}{arg\,max}

\DeclareMathOperator{\diag}{diag}

\DeclareMathOperator{\expec}{E}

\newcommand*{\C}{\mathbb{C}}

\newcommand*{\R}{\mathbb{R}}

\newcommand{\herm}{{\operatorname{H}}}
\newcommand{\tp}{{\operatorname{T}}}

\definecolor{myblue}{RGB}{30, 100, 200}

\newlength{\leftstackrelawd}
\newlength{\leftstackrelbwd}
\def\leftstackrel#1#2{\settowidth{\leftstackrelawd}%
	{${{}^{#1}}$}\settowidth{\leftstackrelbwd}{$#2$}%
	\addtolength{\leftstackrelawd}{-\leftstackrelbwd}%
	\leavevmode\ifthenelse{\lengthtest{\leftstackrelawd>0pt}}%
	{\kern-.5\leftstackrelawd}{}\mathrel{\mathop{#2}\limits^{#1}}}

\newcommand{\mbC}{\bm{C}}

\newcommand{\mbP}{\bm{P}}
\newcommand{\mbQ}{\bm{Q}}

\newcommand{\mbc}{\bm{c}}

\newcommand{\mbf}{\bm{f}}

\newcommand{\mbh}{\bm{h}}

\newcommand{\mbn}{\bm{n}}

\newcommand{\mbp}{\bm{p}}

\newcommand{\mbv}{\bm{v}}

\newcommand{\mbx}{\bm{x}}
\newcommand{\mby}{\bm{y}}
\newcommand{\mbz}{\bm{z}}

\newcommand{\mbSigma}{{\bm{\Sigma}}}

\newcommand{\mbmu}{{\bm{\mu}}}

\newcommand{\inv}{^{-1}}

\newcommand{\hhat}{\hat{\mbh}}

\usetikzlibrary{intersections,pgfplots.fillbetween,patterns,spy}

\Crefname{figure}{Fig.}{Figs.}
\pgfplotsset{compat=1.15}

\ifdefined\ONECOLUMN

\else

\fi

\newcommand{\lineWidth}{1.0pt}
\newcommand{\markSize}{2.0pt}
\definecolor{ourdarkblue}{RGB}{30, 100, 200}
\definecolor{ourdarkgreen}{RGB}{0, 100, 0}
\definecolor{ourdarkorange}{RGB}{201, 98, 18}
\definecolor{ouryellow}{RGB}{220, 210, 50}

\tikzset{tsVQVAE/.style={mark options={solid}, color=TUMBeamerBlue, line width=\lineWidth, mark=square, mark size=\markSize}}
\tikzset{tsVQDAE/.style={mark options={solid}, color=TUMBeamerGreen, line width=\lineWidth, mark=pentagon, mark size=\markSize}}
\tikzset{tsGMMDFT/.style={mark options={solid}, color=TUMBeamerOrange, line width=\lineWidth, mark=o, mark size=\markSize}}
\tikzset{tsLSEDFT/.style={mark options={solid}, color=TUMBeamerRed, line width=\lineWidth, mark=triangle, mark size=\markSize}}

\newacronym{AWGN}{AWGN}{additive white Gaussian noise}
\newacronym{BLMMSE}{BLMMSE}{Bussgang LMMSE}
\newacronym{BS}{BS}{base station}
\newacronym{CDF}{CDF}{cumulative distribution function}
\newacronym{CNN}{CNN}{convolutional neural network}
\newacronym{CSI}{CSI}{channel state information}
\newacronym{CSIT}{CSIT}{channel state information at the transmitter}
\newacronym{DFT}{DFT}{Discrete Fourier transform}
\newacronym{DL}{DL}{downlink}
\newacronym{DNN}{DNN}{deep neural network}
\newacronym{DoA}{DoA}{direction of arrival}
\newacronym{EM}{EM}{expectation maximization}
\newacronym{FDD}{FDD}{frequency division duplex}
\newacronym{GMM}{GMM}{Gaussian mixture model}
\newacronym{LMMSE}{LMMSE}{linear minimum mean square error}
\newacronym{LOS}{LOS}{line of sight}
\newacronym{LS}{LS}{least squares}
\newacronym{MSE}{MSE}{mean squared error}
\newacronym{MIMO}{MIMO}{multiple-input multiple-output}
\newacronym{MPC}{MPC}{multi-path component}
\newacronym{MT}{MT}{mobile terminal}
\newacronym{NLOS}{NLOS}{non-line of sight}
\newacronym{NN}{NN}{neural network}
\newacronym{O2I}{O2I}{outdoor-to-indoor}
\newacronym{OMP}{OMP}{orthogonal matching pursuit}
\newacronym{PDF}{PDF}{probability density function}
\newacronym{PGA}{PGA}{projected gradient ascent}
\newacronym{PSD}{PSD}{power spectral density}
\newacronym{SNR}{SNR}{signal-to-noise ratio}
\newacronym{TDD}{TDD}{time division duplex}
\newacronym{UL}{UL}{uplink}
\newacronym{ULA}{ULA}{uniform linear array}
\newacronym{URA}{URA}{uniform rectangular array}
\newacronym{UMa}{UMa}{urban macrocell}
\newacronym{nSE}{nSE}{normalized spectral efficiency}
\newacronym{cCDF}{cCDF}{complementary cumulative distribution function}
\newacronym{MU-MIMO}{MU-MIMO}{multi-user MIMO}
\newacronym{MU-MISO}{MU-MISO}{multi-user MISO}
\newacronym{BD}{BD}{block diagonalization}
\newacronym{RBD}{RBD}{regularized block diagonalization}
\newacronym{RCI}{RCI}{regularized channel inversion}
\newacronym{WMMSE}{WMMSE}{weighted minimum mean square error}
\newacronym{SWMMSE}{SWMMSE}{stochastic WMMSE}
\newacronym{SVD}{SVD}{singular value decomposition}
\newacronym{SR}{SR}{sum-rate}
\newacronym{CME}{CME}{conditional mean estimator}
\newacronym{ML}{ML}{machine learning}
\newacronym{FLOPS}{FLOPS}{floating-point operations}
\newacronym{OFDM}{OFDM}{orthogonal frequency-division multiplexing}
\newacronym{LTE}{LTE}{Long Term Evolution}
\newacronym{GPS}{GPS}{Global Positioning System}
\newacronym{UMi}{UMi}{urban microcell}
\newacronym{VQ-VAE}{VQ-VAE}{vector quantized-variational autoencoder}
\newacronym{AE}{AE}{autoencoder}
\newacronym{3GPP}{3GPP}{3rd Generation Partnership Project}

\newcommand{\Nv}{N_{\mathrm{v}}}
\newcommand{\Nh}{N_{\mathrm{h}}}

\begin{document}

\title{Feedback Design with VQ-VAE for \\Robust Precoding in Multi-User FDD Systems}


\author{Nurettin~Turan~\IEEEmembership{Graduate Student Member,~IEEE,} Michael Baur~\IEEEmembership{Graduate Student Member,~IEEE,} Jianqing Li,\\ and Wolfgang~Utschick~\IEEEmembership{Fellow,~IEEE}

\thanks{The authors are with the TUM School of Computation, Information and Technology, Technische Universität München, 80333 Munich, Germany \textit{Corresponding author: Nurettin Turan} (e-mail: nurettin.turan@tum.de).\\
N. Turan and M. Baur contributed equally to this work.\\
The authors acknowledge the financial support by the Federal Ministry of
Education and Research of Germany in the program of ``Souver\"an. Digital.
Vernetzt.''. Joint project 6G-life, project identification number: 16KISK002.\\
This work is funded by the Bavarian Ministry of Economic Affairs, Regional
Development, and Energy as part of the project 6G Future Lab Bavaria.
}
}


\maketitle

\begin{abstract}
In this letter, we propose a \ac{VQ-VAE}-based feedback scheme for robust precoder design in multi-user \ac{FDD} systems. 
We demonstrate how the \ac{VQ-VAE} can be tailored to specific propagation environments, focusing on systems with low pilot overhead, which is crucial in massive \ac{MIMO}. 
Extensive simulations with real-world measurement data show that our proposed feedback scheme outperforms state-of-the-art \ac{AE}-based compression schemes and conventional \ac{DFT} codebook-based schemes. 
These improvements enable the deployment of systems with fewer feedback bits or pilots.

\end{abstract}


\vspace{-0.5em}
\section{Introduction}

\vspace{-1em}
\begin{tikzpicture}[remember picture,overlay]
\node[anchor=south,yshift=14pt,xshift=-134pt] at (current page.south) {{\parbox{\dimexpr\columnwidth-\fboxsep-\fboxrule\relax}{\footnotesize \quad \copyright This work has been submitted to the IEEE for possible publication. Copyright may be transferred without notice, after which this version may no longer be accessible.}}};
\end{tikzpicture}%

In the next generation of cellular communications systems (6G), the \ac{BS} will have the capability to adapt to dynamic channel conditions.
In massive \ac{MIMO} \ac{FDD} systems, this adaption is maintained by regular \ac{CSI} feedback from the \acp{MT} due to the absence of channel reciprocity~\cite{Love}.

In this regard, conventional schemes entail estimating the \ac{DL} channel at the \acp{MT} and subsequently determining the feedback information using codebook-based schemes~\cite{Love, KaKoGeKn09}. 
In scenarios with spatial correlation due to specific antenna geometries, \ac{DFT}-based codebooks are well-established and are part of \ac{3GPP} specifications \cite{LiSuZeZhYuXiXu13, 3GPP_5G}.
Alternatively, the feedback information can be obtained with state-of-the-art \ac{ML}-based \ac{CSI} compression schemes, e.g., \cite{WeShJi18, GuWeJiLi20, RiNeJoClUt23, LePaKwLeCh24}.
In the seminal work \cite{WeShJi18}, \acp{AE} were utilized for \ac{CSI} feedback, where the encoder is used to compress at the \ac{MT} side, and the decoder reconstructs at the \ac{BS} side. 
Building on this concept, \cite{GuWeJiLi20} introduced quantization in the \ac{AE}'s latent space.
In \cite{RiNeJoClUt23}, \acp{VQ-VAE} were adopted to learn the quantization operation using an embedding in the latent space \cite{OoVinKa17}.
Moreover, \cite{RiUt21, LePaKwLeCh24} explored the use of \acp{AE} for reconstructing channels from noisy input data, leveraging the denoising capabilities of \acp{AE} \cite{ViLaBeMa08}.

However, in massive \ac{MIMO} \ac{FDD} systems, where the \ac{BS} is typically equipped with a high number of antennas, as many pilots as transmit antennas are required to be sent from the \ac{BS} to the \acp{MT} to fully illuminate the channel.
This results in a significant pilot overhead, which can be prohibitive \cite{BjLaMa16}.
In this work, in addition to noisy input data, system setups with fewer pilots than transmit antennas are considered, rendering the instantaneous reconstruction of the \ac{CSI} particularly challenging.
To address this, we propose a \ac{VQ-VAE}-based feedback scheme for robust precoder design.
The contributions of this work can be summarized as follows:

\begin{itemize}
    \item We propose to utilize VQ-VAEs in combination with a state-of-the-art stochastic algorithm for robust precoder design in multi-user systems.
    To enhance the \ac{VQ-VAE} training, we suggest adapting the loss function and the architecture based on model-based insights.
    Specifically, the covariance matrix at the VQ-VAE's decoder output is constrained according to the array geometry at the \ac{BS} side, a scalar embedding is employed, and the \ac{VQ-VAE} input undergoes a pre-transformation.
    \item Due to the applied training procedure, the proposed scheme provides easy adaption to any desired \ac{SNR} and supports multi-user operation for any number of \acp{MT} without requiring retraining.
    \item Extensive simulations using real-world measurement data demonstrate that the proposed feedback scheme for robust precoder design outperforms state-of-the-art \ac{AE}-based compression schemes trained for instantaneous reconstruction, as well as conventional \ac{DFT} codebook-based schemes. 
    The performance improvements achieved with the proposed scheme enable the deployment of systems with fewer feedback bits or pilots.
\end{itemize}

\section{System Model and Channel Data}
\label{sec:system_channel_model}

\subsection{Data Transmission Phase}

We consider the \ac{DL} of a single-cell multi-user system, where the \ac{BS} equipped with $N$ transmit antennas serves $J$ single-antenna \acp{MT}.
Linear precoding is adopted, with the precoded \ac{DL} data vector given by $\mbx = \sum_{j=1}^{J}\mbv_js_j$, where $s_j$ is the transmit signal for \ac{MT} $j$, $\expec[s_j]=0$, $\expec[\left|s_j\right|^2]=1$, and $\mbv_j \in \C^{N}$ is the precoding vector of \ac{MT} $j$.
The precoders satisfy the transmit power constraint $\sum_{j=1}^J \operatorname{tr}(\mbv_j^\herm \mbv_j) = \rho$. 
The sum-rate is expressed as
\begin{equation} 
\label{eq:inst_sumrate}
    R = \sum_{j=1}^J \log_2 \Bigg(1 + \dfrac{\left|\mbh_j^\tp\mbv_j\right|^2}{ \sum_{j^\prime \neq j} \left|\mbh_j^\tp\mbv_{j^\prime}\right|^2 + \sigma_n^2}\Bigg),
\end{equation}
where $\mbh_j\in \C^{N}$ denotes the channel of \ac{MT} $j$ and $\sigma_n^2$ is the noise variance.
In the considered \ac{FDD} system, the \ac{BS} designs the precoders $\mbv_j$ based on each \ac{MT}'s feedback information, which is encoded by $B$ bits.

\subsection{Pilot Transmission Phase}

Before data transmission, ${n_\mathrm{p}}$ orthogonal pilots are broadcasted to all \acp{MT}, allowing each \ac{MT} to infer its feedback information.
During the pilot transmission phase, the received signal $\mby_j \in \C^{{n_\mathrm{p}}} $ of each \ac{MT} is given by
\begin{equation} \label{eq:noisy_obs}
    \mby_j = \mbP \mbh_j + \mbn_j
\end{equation}
where $\mbn_j \sim \mathcal{N}_\C(\mathbf{0}, \mbSigma)$ represents the \ac{AWGN} with $\mbSigma = \sigma_n^2 \mathbf{I}_{{n_\mathrm{p}}}$.
We utilize a $2$D-DFT (sub)matrix as the pilot matrix $\mbP$, which is motivated by the \ac{URA} at the BS, cf. \cite{TsZhWa18}.
To meet the power constraint, each column $\mbp_\ell$ of $\mbP^\tp$ is normalized such that $\|\mbp_\ell\|_2^2=\rho$ for all $\ell \in \{1,2, \dots, {n_\mathrm{p}}\} $.
We consider systems with fewer pilots than transmit antennas, i.e., ${n_\mathrm{p}} < N$.

\subsection{Real-World Channel Data} \label{sec:data_generation}

\begin{figure}
    \centering
    \includegraphics[]{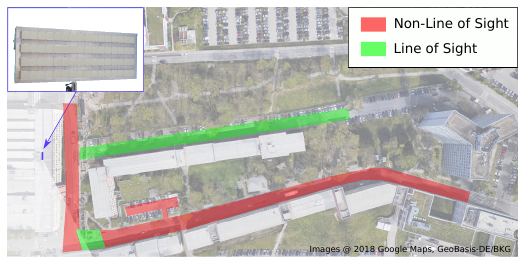}
    \caption{Measurement campus in Stuttgart, Germany \cite{HeDeWeKoUt19}.}
    \label{fig:meas_campaign}
\end{figure}

The measurement campaign took place at the Nokia campus in Stuttgart, Germany, in 2017.
As shown in \Cref{fig:meas_campaign}, the \ac{BS} antenna, with a $\SI{10}{\degree}$ down-tilt, was installed on a rooftop approximately $\SI{20}{m}$ above the ground.
The antenna array geometry was tailored to the \ac{UMi} propagation scenario.
Thus, the \ac{BS} comprises a \ac{URA} with $\Nv=4$ vertical and $\Nh=16$ horizontal single polarized patch antennas, totaling $N=64$ elements.
The horizontal spacing was $\lambda/2$, and the vertical spacing was~$\lambda$, where $\lambda$ is the wavelength. 
The carrier frequency used was $\SI{2.18}{\giga\hertz}$. 
The single monopole receive antenna, representing the \acp{MT}, was mounted on a vehicle at a height of $\SI{1.5}{m}$, which moved at a maximum speed of $\SI{25}{kmph}$. 
Further details can be found in \cite{HeDeWeKoUt19}.

\section{Proposed VQ-VAE-based Feedback Scheme for Robust Precoder Design} \label{sec:propscheme}

In general, the \ac{VQ-VAE}'s loss function is given by, cf. \cite{OoVinKa17},
\begin{equation} \label{eq:vqvae_loss}
    \mathcal{L}_{\mathrm{VQ-VAE}} =  \mathcal{L}_{\mathrm{rec}} + \| \mathrm{sg}(\mbz_j) - \mbf_j\|_2^2 + \beta \| \mbz_j - \mathrm{sg}(\mbf_j)\|^2_2,
\end{equation}
where $\mathcal{L}_{\mathrm{rec}}$ denotes the reconstruction loss and $\mathrm{sg}(\cdot)$  denotes the stop gradient operator, which treats its argument as a constant.
The second term in \eqref{eq:vqvae_loss} is the dictionary learning term, which ensures that the embedding $\mathcal{E}$ is learned such that the output after quantization, i.e., $\mbf_j$ is as close as possible to the unquantized latent representation $\mbz_j$.
The third term in \eqref{eq:vqvae_loss} is the commitment loss, encouraging the encoder to commit to the embedding $\mathcal{E}$ and ensuring that the encoder output does not grow unbounded.
The balancing hyper-parameter $\beta=0.25$ is introduced to facilitate a good reconstruction by weighting the commitment loss lower and thereby enhancing the impact of the dictionary learning \cite{OoVinKa17, RiNeJoClUt23}.
To maintain a continuous gradient flow between the encoder and decoder, despite the discontinuity introduced by quantization, the gradients from the decoder input to the encoder output are copied using $\mbf_j = \mbz_j + \mathrm{sg}(\mbf_j-\mbz_j)$.
In summary, the encoder is trained using the first and third terms of the loss \eqref{eq:vqvae_loss}, the embedding $\mathcal{E}$ is learned through the second term, and the decoder is trained with the first term.

\begin{figure}
    \centering
    \includegraphics[scale=0.73]{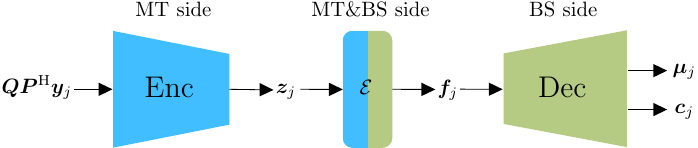}
    \caption{Structure of the proposed \ac{VQ-VAE} for robust precoder design.}
    \label{fig:VQVAE_struct}
\end{figure}

The \ac{VQ-VAE} can be tailored to the \ac{BS} environment characteristics to enable robust precoder design as follows.
We require the \ac{VQ-VAE} to output a mean vector $\mbmu_j$, and another vector $\mbc_j$ parametrizing a covariance matrix as approximate statistical information about the channel $\mbh_j$ (see \Cref{fig:VQVAE_struct}).
Since we have a \ac{URA} deployed at the \ac{BS} side, a reasonable choice is to constrain the covariance matrices to be block-Toeplitz matrices with Toeplitz blocks:
\begin{equation}
    \mbC_j = \mbQ^\herm \diag(\mbc_{j}) \mbQ
\end{equation}
with  $\mbQ = \mbQ_{\Nv} \otimes \mbQ_{\Nh}$, where $\mbQ_T$ (with $T \in \{\Nv, \Nh\}$) contains the first $T$ columns of a $2T\times 2T$ \ac{DFT} matrix and $\mbc_j \in \R_{+}^{4N}$, see~\cite{BaBoTuUt24}.
Consequently, the covariance matrices are fully characterized by the output vectors $\mbc_{j}$.
This structural constraint acts as a regularization and helps to decrease the model parameters since
the decoder output vector $\mbc_j$ fully determines the covariance matrix $\mbC_j$.

We adopt a similar encoder and decoder architecture as in~\cite{BaBoTuUt24}, where we adapt the input and latent dimension accordingly.
We also adopt an embedding $\mathcal{E}$ consisting of in total $C$ scalar entries, i.e., $\mathcal{E}=\{e_1, e_2, \dots, e_C\}$, keeping in mind that also the embedding needs to be shared with the \acp{MT}.
For a fixed latent space dimension $N_\mathrm{L}$, the $i$-th entry of the feedback vector $\mbf_j$ is obtained by using the embedding $\mathcal{E}$ as
\begin{equation}
    f_{j,i} = \min_{e_c \in \mathcal{E}} | e_c -  z_{j,i}|,
\end{equation}
where $z_{j,i}$ denotes the $i$-th entry of the unquantized latent representation $\mbz_j$, and $i=1,\dots,N_\mathrm{L}$.
In this way, the feedback vector $\mbf_j$ can be fully described by $B=N_\mathrm{L} \log_2 C$ bits.
In principle, sub-chunks of $\mbz_j$ could be quantized with vectors instead of scalars, see \cite{OoVinKa17}.
However, in our simulations, we did not observe any gains with an embedding consisting of vectors of relatively low dimensions instead of scalars.
With vectors of too large dimensions, the training procedure is even infeasible if a feedback size of tens to hundreds of bits is desired. 

The observation in \eqref{eq:noisy_obs} is first multiplied with the pilot matrix to obtain a coarse estimate.
Subsequently, we apply $\mbQ$ as a transformation, which is inspired by \cite{NeWiUt18} and can be interpreted as an angular domain transformation, cf., e.g., \cite{WeShJi18}.
Accordingly, the input to the \ac{VQ-VAE} is given by $\mbQ \mbP^\herm \mby_j$.

Taking into account the adaptions to tailor the \ac{VQ-VAE} for robust precoder design for a specific \ac{BS} environment, the overall loss function is the loss from \eqref{eq:vqvae_loss} using the embedding $\mathcal{E}=\{e_1, e_2, \dots, e_C\}$ with the reconstruction loss
\begin{equation}
    \mathcal{L}_{\mathrm{rec}} = 
    \log \det (\pi \mbC_j) + (\mbh_j -\mbmu_j)^\herm  \mbC_j\inv
    (\mbh_j - \mbmu_j).
\end{equation}
Multi-user operation for any desired number of \acp{MT} $J$ is supported since the \ac{VQ-VAE} is trained to allow the inference of statistical information about the channel of any \ac{MT} within the \ac{BS} cell environment.
To enable operation at different \ac{SNR} levels, we train the \ac{VQ-VAE} for an \ac{SNR} range from $0$ to $\SI{20}{dB}$.
In particular, for the offline training, we utilize the training set $\{\mbh^{(m)}\}_{m=1}^M$ and generate correspondingly encoder inputs $\mbQ \mbP^\herm \mby^{(m)}$ according to \eqref{eq:noisy_obs} where we randomly draw a noise variance to obtain an \ac{SNR} from the range specified above.

In \Cref{fig:VQVAE_struct}, the \ac{VQ-VAE} with the proposed adaptions is depicted.
As discussed before, knowledge of the encoder and the embedding is required to enable the \acp{MT} to infer their feedback information.
In the online phase, given the observation $\mby_j$ at each \ac{MT} $j$, the observation is preprocessed, passed through the encoder to obtain the unquantized latent variable $\mbz_j$, and finally quantized using the embedding $\mathcal{E}$ to obtain the feedback information vector $\mbf_j$. 
Each \ac{MT} $j$ sends the feedback information encoded with $N_\mathrm{L} \log_2 C$ bits back to the \ac{BS}.
At the \ac{BS} side, the acquired statistical information about each \ac{MT} is used in the following way to design precoders.

We utilize the sample generation capabilities of \acp{VQ-VAE}, treating the \ac{MT} channels as random variables, and apply a stochastic version of the well-known iterative \ac{WMMSE} algorithm~\cite{ShRaLuHe11}, specifically the \ac{SWMMSE} algorithm from \cite{RaSaLu16} for robust precoder design.
In each iteration step, the \ac{SWMMSE} requires samples that follow the channel distribution of each \ac{MT}.
Therefore, based on each \ac{MT}'s feedback information, we propose generating samples using the \ac{VQ-VAE} as
\begin{equation}
     \tilde{\bm{h}}_j \sim \mathcal{N}_{\mathbb{C}}(\bm{\mu}_{j}, \bm{C}_{j}),
\end{equation}
resembling samples from the channel distribution of \ac{MT} $j$.
The \ac{BS} uses the generated samples in each iteration step of the \ac{SWMMSE} algorithm to jointly design the precoders for all \acp{MT}.
A summary of the proposed \ac{VQ-VAE}-based feedback scheme for robust precoder design is provided in \Cref{alg:vqvae_rpd}.

\begin{algorithm}[t]
\captionsetup{font=footnotesize}
\caption{VQ-VAE-based Feedback Scheme for Robust Precoder Design}
\label{alg:vqvae_rpd}
    \begin{algorithmic}[1] \footnotesize
    \REQUIRE Offline trained VQ-VAE where encoder ($\mathrm{Enc}$) and embedding $\mathcal{E}$ are available at the \acp{MT} and decoder ($\mathrm{Dec}$) and embedding $\mathcal{E}$ are available at the \ac{BS}.
    
    \hspace{-0.6cm}{\textbf{Feedback Inference at the MTs}}
    
    \STATE Infer unqunatized latent $\mbz_j$ from prepocessed observation at each MT $j$
    
    $\mbz_j = \mathrm{Enc}(\mbQ\mbP^\herm\mby_j) \ \forall j$

    \STATE Obtain quantized latent representation $\mbf_j$ at each MT $j$

    $ f_{j,i} = \min_{e_c \in \mathcal{E}} |e_c -  z_{j,i}| \ \text{with} \ i=1,\dots,N_\mathrm{L} \ \forall j$

    \STATE Each MT $j$ feeds back $\mbf_j$ encoded by $B=N_\mathrm{L}\log_2C$ bits
    
    \hspace{-0.6cm}{\textbf{Robust Precoder Design at the BS}}

    \STATE Obtain statistical description of each MT $j$ 
    
    $(\mbmu_j, \mbc_j) = \mathrm{Dec}(\mbf_j) \ \forall j$

    \STATE Compute covariance matrix $\mbC_j$ for each MT $j$

    $\mbC_j = \mbQ^\herm \diag(\mbc_{j}) \mbQ \ \forall j$

    \STATE Provide statistics of each MT $j$ for sampling in the SWMMSE algorithm and obtain precoders
    
    $\{\mbv_j\}_j^J \gets \text{SWMMSE}( \{\tilde{\bm{h}}_j \sim \mathcal{N}_{\mathbb{C}}(\bm{\mu}_{j}, \bm{C}_{j}) \}_{j=1}^J )$
    \end{algorithmic}
\end{algorithm}

\section{Baseline Feedback Schemes for Precoding}

\label{sec:baselines}
\subsection{State-of-the-art AE and VQ-VAE-based Feedback Schemes}
\label{sec:vqvaei_baseline}
In state-of-the-art feedback schemes involving \acp{AE} or \acp{VQ-VAE}, a reconstruction $\bar{\mbh}_j$ of the channel of \ac{MT} $j$ is obtained at the decoder output, cf. \cite{WeShJi18, RiNeJoClUt23}.
To facilitate a fair comparison, we utilize the same \ac{VQ-VAE} architecture, encoder input pre-processing, and an embedding $\mathcal{E}$ of the same size.
The main adaption is thus given by the decoder output, where the output covariance is fixed to be the identity and only a mean vector, denoted by $\bar{\mbh}_j$, is inferred.
Accordingly, the overall loss function which aims to reconstruct the instantaneous channel is the loss from \eqref{eq:vqvae_loss} with the \ac{MSE} as reconstruction loss 
\begin{equation}
    \mathcal{L}_{\mathrm{rec}} =  \|\mbh_j - \bar{\mbh}_j\|_2^2.
\end{equation}
In the case of the regular \ac{AE}, the embedding is omitted, and the overall loss degenerates to the \ac{MSE} reconstruction loss. 
Also, in these cases, multi-user operation for any desired number of \acp{MT} $J$ is supported, and we train for the same \ac{SNR} range to facilitate operation for different \ac{SNR} levels.
Since the information about each \ac{MT} $j$ is a reconstruction $\bar{\mbh}_j$, we apply the iterative \ac{WMMSE} algorithm for precoder design.
Specifically, we use \cite[Algorithm~1]{HuCaYuQiXuYuDi21}.

\subsection{Conventional DFT Codebook-based Feedback Scheme}
\label{sec:dft_feedback}

In conventional codebook-based feedback schemes, each \ac{MT} first computes a channel estimate $\hat{\mbh}_j$ and then determines the channel directional information as (see, e.g.,~\cite{KaKoGeKn09}):
\begin{equation} \label{eq:dft_idx}
    k^\star_j = \argmax_{k } |\mbc_k^\herm \hat{\mbh}_j|
\end{equation}
where $\mbc_k \in \mathcal{C}$, with the codebook $\mathcal{C} = \{\mbc_1, \dots, \mbc_{K_\text{Dir}} \}$ of cardinality $|\mathcal{C}| = K_\text{Dir} = 2^{B_\text{Dir}}$ with $B_\text{Dir}$ bits.
Since we have a \ac{URA} at the \ac{BS}, a 2D-\ac{DFT} codebook is used, which is constructed by the Kronecker product of two under-/oversampled \ac{DFT} matrices~\cite{LiSuZeZhYuXiXu13}.
The under-/oversampling factors are chosen to satisfy $|\mathcal{C}| = K_\text{Dir} = 2^{B_\text{Dir}}$ for a given $B_\text{Dir}$.
Furthermore, the channel magnitude information $m_j$ is assumed to be the norm of the estimated channel $\hat{\mbh}_j$ and is encoded using $B_\text{Mag}$ bits \cite{KaKoGeKn09}.
Thus, the overall feedback information is encoded using $B=B_\text{Dir} + B_\text{Mag}$ bits.
Based on the feedback from each \ac{MT}, the \ac{BS} represents the channel of each \ac{MT} as \cite{KaKoGeKn09},
\begin{equation} \label{eq:dft_approach}
    \hat{\mbh}^{(q)}_j = m_j \mbc_{k^\star_j},
\end{equation}
and then utilizes the iterative \ac{WMMSE} \cite[Algorithm~1]{HuCaYuQiXuYuDi21} to jointly design the precoders.

\section{Simulation Results} \label{sec:sim_results}

\begin{figure}[tb]
    \centering
    \includegraphics[]{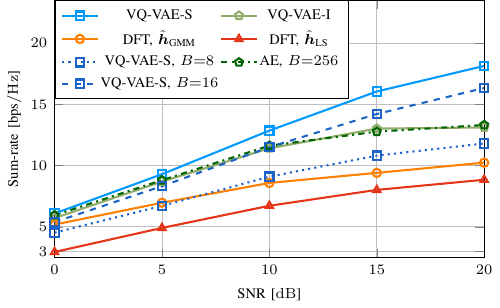}
    \caption{The sum-rate over the \ac{SNR} for a system with $B=40$ feedback bits, $J=8$ \acp{MT}, and $n_{\mathrm{p}}=8$ pilots.}
    \label{fig:fig_40bits_np8_oversnr_Xbits}
    \vspace{-0.5cm}
\end{figure}

We use $420{,}000$ measurement data samples from the measurement campaign detailed in \Cref{sec:data_generation}.
Specifically, $M= 400{,}000$ samples are used for training the data-based methods, $10{,}000$ for validation, and $10{,}000$ for evaluation.
The data samples are normalized such that \( \expec[\|\mbh\|^2] = N \). 
Additionally, we set $\rho=1$, allowing the \ac{SNR} to be defined as~\( \frac{1}{\sigma_n^2} \).
The sum-rate averaged over $500$ multi-user constellations is employed as the performance metric, where $J$ \acp{MT} are randomly selected from our evaluation set for each constellation.
For the \ac{SWMMSE} as well as the iterative \ac{WMMSE}, we set the maximum number of iterations to $I_{\max}=300$.
For clarity, we omit the index $j$ in the subsequent descriptions.

In the following discussion, the proposed \ac{VQ-VAE}-based feedback scheme from \Cref{sec:propscheme}, which utilizes statistical information for precoder design, is denoted by ``VQ-VAE-S.''
The baseline \ac{VQ-VAE} feedback scheme from \Cref{sec:vqvaei_baseline}, which uses the instantaneous reconstructions, is denoted by ``VQ-VAE-I,'' and the regular \ac{AE} with no quantization in the latent space by ``AE.''
As outlined in \Cref{sec:dft_feedback}, the \ac{DFT} codebook-based feedback scheme requires channel estimation.
We consider two estimators: the \ac{GMM}-based channel estimator \( \hhat_{\text{GMM}} \) with $64$ components \cite{KoFeTuUt21J} and the conventional \ac{LS} estimator $\hhat_{\text{LS}}$.
Cases where channel estimation is performed at each \ac{MT} and the feedback information is then determined using the \ac{DFT} codebook-based scheme are denoted as ``DFT, \{$\hhat_{\text{GMM}}$, $\hhat_{\text{LS}}$\}.''

In the subsequent simulations, we fix the feedback budget to $B=40$ bits, if not mentioned otherwise.
This is motivated by a typical compression ratio for the regular \ac{AE} with $N_\mathrm{L}=8$, cf. \cite{WeShJi18}, where each latent entry is described by $32$ bits (single precision), yielding $8\cdot32=256$ bits in total (``\ac{AE}, $B{=}256$'').
We observed, by introducing the embedding with as few as $C=32$ (i.e. $5$ bits per entry), the same performance was achieved with the ``VQ-VAE-I,'' see \Cref{fig:fig_40bits_np8_oversnr_Xbits}.
For the \ac{VQ-VAE} based methods ``VQ-VAE-S'' and ``VQ-VAE-I,'' we thus fix $N_\mathrm{L}=8$ and set $C=32$ ($N_\mathrm{L} \log_2 C = 8 \cdot 5 = 40$ bits).
For the \ac{DFT} codebook-based scheme we allocate $B_\text{Dir} = 8$ bits for the directional information and $B_\text{Mag}=32$ bits for the magnitude (single precision), ensuring $B_\text{Dir} + B_\text{Mag} = 40$ bits.
While different bit allocations for $B_\text{Dir}$ and $B_\text{Mag}$ are possible to achieve a total of $B=40$ bits, the chosen allocation maintains reasonable complexity for determining channel directional information. 
Additionally, the ``VQ-VAE-I'' method inherently aims for a more efficient bit allocation for improved instantaneous reconstruction compared to the conventional \ac{DFT} codebook-based scheme.

\begin{figure}[tb]
    \centering
    \includegraphics[]{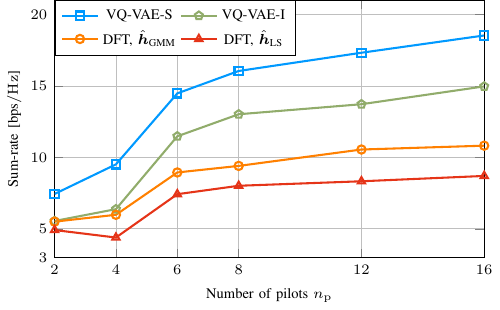}
    \caption{The sum-rate over the number of pilots $n_\mathrm{p}$ for a system with $J=8$ \acp{MT}, $B=40$ feedback bits, and $\text{SNR}=\SI{15}{dB}$.}
    \label{fig:fig_40bits_snr15_overpilots}
    \vspace{-0.5cm}
\end{figure}

In \Cref{fig:fig_40bits_np8_oversnr_Xbits}, we have $J=8$ \acp{MT} and $n_{\mathrm{p}}=8$ pilots and depict the sum-rate over the \ac{SNR}. 
The proposed scheme ``VQ-VAE-S'' performs best, followed by the ``VQ-VAE-I.''
The conventional approaches ``DFT, \{$\hhat_{\text{GMM}}$, $\hhat_{\text{LS}}$\}'' perform poorly, especially at high \ac{SNR} levels.
Despite improvements with the \ac{GMM} estimator over the \ac{LS} estimator, the performance remains poor due to significant channel estimation errors.
Our proposed scheme with even a reduced feedback bit overhead ``VQ-VAE-S, $B=16$'' ($N_\mathrm{L}=8$, $C=4$) performs comparable to the ``VQ-VAE-I'' approach with $B=40$ bits in the low \ac{SNR} regime, and is superior for high \ac{SNR} levels.
The proposed scheme with only $8$ feedback bits denoted by ``VQ-VAE-S, $B=8$'' ($N_\mathrm{L}=8$, $C=2$) significantly outperforms ``DFT, $\hhat_{\text{LS}}$'' and performs comparable to ``DFT, $\hhat_{\text{GMM}}$'' in the low \ac{SNR} regime, and is superior for high \ac{SNR} levels.
Accordingly, the performance gains from the proposed \ac{VQ-VAE}-based robust precoder design scheme allow for deploying systems with reduced feedback overhead while maintaining performance.

In \Cref{fig:fig_40bits_snr15_overpilots}, we examine the impact of the number of pilots $n_\mathrm{p}$ on the performance with $J=8$ \acp{MT} for a fixed $\text{SNR}$ of $\SI{15}{dB}$.
The proposed scheme ``VQ-VAE-S'' performs best for all considered numbers of pilots $n_\mathrm{p}$.
With only $n_\mathrm{p}=6$ pilots, the ``VQ-VAE-S'' approach performs almost as well as the ``VQ-VAE-I'' baseline with $n_\mathrm{p}=16$ pilots.
Additionally, with $n_\mathrm{p}=4$ pilots, the ``VQ-VAE-S'' approach performs almost as well as the ``DFT, $\hhat_{\text{GMM}}$'' approach and outperforms ``DFT, $\hhat_{\text{LS}}$'' each with $n_\mathrm{p}=16$ pilots.
This analysis reveals that systems with lower pilot overhead can be deployed without sacrificing performance compared to the baseline schemes.

\begin{figure}[tb]
    \centering
    \includegraphics[]{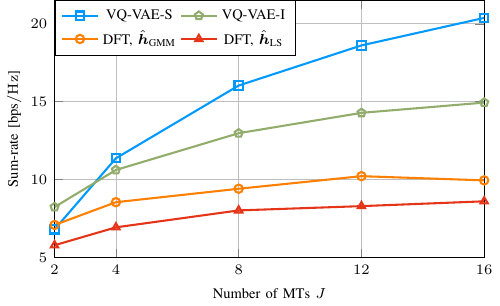}
    \caption{The sum-rate over the number of \acp{MT} $J$ for a system with $B=40$ feedback bits, $n_\mathrm{p}=8$ pilots, and $\text{SNR}=\SI{15}{dB}$.}
    \label{fig:fig_40bits_snr15_np8_overuser}
    \vspace{-0.3cm}
\end{figure}

In \Cref{fig:fig_40bits_snr15_np8_overuser}, we set $n_{\mathrm{p}}=8$ pilots, and $\text{SNR}=\SI{15}{dB}$ and vary the number $J$ of served \acp{MT}.
It can be seen that the \ac{VQ-VAE}-based feedback scheme ``VQ-VAE-S'' is superior as compared to the baselines for all numbers of \acp{MT} except for a setup with only two \acp{MT}.
Remarkably, the sum-rate of the ``VQ-VAE-S'' approach steadily increases with an increasing number of \acp{MT}.
The remaining approaches exhibit saturation or even slightly degrade with an increasing number of \acp{MT}.
This can be reasoned by the fact that with an increasing number of \acp{MT}, resolving the interference present in the scenario becomes more difficult, particularly when restricted to a directional codebook of finite size.
In contrast, with the proposed robust approach ``VQ-VAE-S,'' due to the involved sampling procedure (see \Cref{alg:vqvae_rpd}), a different representative interference scenario is provided to and exploited by the \ac{SWMMSE} to design the precoders.
For a system with only two \acp{MT}, the interference is less severe, and thus, the ``VQ-VAE-I'' aiming for an instantaneous reconstruction is beneficial.

\begin{figure}[tb]
    \centering
    \includegraphics[]{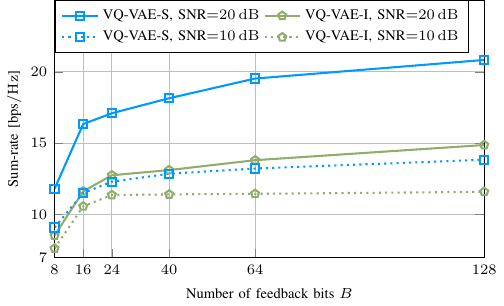}
    \caption{The sum-rate over the number of feedback bits $B$, for a system with $J=8$ \acp{MT}, $n_\mathrm{p}=8$ pilots, and $\text{SNR} \in \{\SI{10}{dB}, \SI{20}{dB}\}$.}
    \label{fig:fig_Xbits_np8_J8_snr10_20}
    \vspace{-0.4cm}
\end{figure}

Lastly, in \Cref{fig:fig_Xbits_np8_J8_snr10_20}, we set $J=8$ \acp{MT}, $n_{\mathrm{p}}=8$ pilots, and $\text{SNR}=\SI{15}{dB}$ and investigate the impact of the number $B$ of feedback bits on the system performance for the two \ac{SNR} levels $\SI{10}{dB}$ and $\SI{20}{dB}$.
For $B\leq40$, we adopt $N_\mathrm{L}=8$ and correspondingly $C\in\{2,4,8,32\}$ (i.e. $\{1,2,3,5\}$ bits per entry).
For $B\geq64$, we adopt $N_\mathrm{L}=32$ and $C\in\{4,16\}$ (i.e. $\{2,4\}$ bits per entry).
The proposed scheme ``VQ-VAE-S'' outperforms the baseline ``VQ-VAE-I'' for all feedback bit configurations by a large margin, especially for $\text{SNR}=\SI{20}{dB}$.
Once again, the results indicate that the proposed scheme ``VQ-VAE-S'' can achieve superior performance than the baseline approach ``VQ-VAE-I'' with fewer bits, reducing both feedback overhead and processing requirements at the \acp{MT}.

\section{Conclusion}

In this letter, we proposed a \ac{VQ-VAE}-based feedback scheme for robust precoder design in multi-user \ac{FDD} systems.
The \ac{VQ-VAE} can be tailored to the characteristics of the propagation environment and outperformed conventional \ac{DFT} codebook-based and state-of-the-art \ac{AE}-based feedback schemes in systems characterized by low pilot overhead.
Future work can enhance the proposed \ac{VQ-VAE}-based feedback scheme by embedding it into an end-to-end system that learns both the pilot matrix and the precoder optimization.

\bibliographystyle{IEEEtran}
\bibliography{IEEEabrv,biblio}
\end{document}